\documentclass{article}
\usepackage{stmaryrd, graphicx, times} 
\usepackage{amsmath}
\usepackage{amscd}
\usepackage{amsthm}
\usepackage{amssymb} \usepackage{latexsym}
\usepackage{eufrak}
\usepackage{euscript}
\usepackage{epsfig}
\usepackage{graphics}
\usepackage{array}
\usepackage{enumerate}

\newcommand{\bel}[1]{\begin{equation}\label{#1}}

\newcommand{\be}{\begin{equation}}

\newcommand{\ba}{\begin{eqnarray}}
\newcommand{\ea}{\end{eqnarray}}
\newcommand{\rf}[1]{(\ref{#1})}
\newcommand{\bi}{\bibitem}
\newcommand{\qe}{\end{equation}}

\begin{document}
\title{Laplacian Spectrum and Protein-Protein Interaction Networks}
\author{Anirban Banerjee, J\"urgen Jost\footnote{Max Planck Institute for Mathematics in the
  Sciences, Inselstr.22, 04103 Leipzig, Germany, banerjee@mis.mpg.de, jost@mis.mpg.de}}

\maketitle
\begin{abstract}
From the spectral plot of the (normalized) graph Laplacian,
   the essential qualitative properties of a
  network  can be simultaneously deduced. Given a class of empirical networks, 
  reconstruction schemes for elucidating the evolutionary dynamics
  leading to those particular data  can then be developed. This method is exemplified for
  protein-protein interaction networks. Traces of their  evolutionary history of
  duplication and divergence processes are identified. In particular,
  we can identify typical specific features that robustly distinguish
  protein-protein interaction networks from other classes of networks,
  in spite of possible statistical fluctuations of the underlying data.
\end{abstract}

\section{Introduction}
In recent years, many studies have investigated certain important
 parameters for empirical networks,  such as degree distribution,
 average path length, diameter, betweenness centrality, transitivity
 or clustering coefficient etc. Such studies could identify certain
 rather universal features valid for networks across a wide range of
 disciplines, like scalefree degree distributions. Conversely, on this
 basis, often algorithms could be developed that, perhaps after
 fitting certain free parameters, could construct networks with the
 same qualitative properties and values for such variables. \\
Here, we look at an essentially complete set of graph variables, given
by the spectrum of its normalized Laplacian. On this basis, we can
then develop algorithms that construct networks with all the essential
qualitative
properties as the ones in a given data set. For
biological networks, we can thereby retrace the regularities in their
evolutionary history. Here, we demonstrate this principle and apply
this method for
protein-protein interaction networks (PPIN for short). We detect indications of an
evolutionary of duplication and divergence, as argued in \cite{Wagner2001,HuynenBork1998}. \\
This approach then also sheds light on a somewhat different issue, namely which
features and properties are distinctive for networks from particular
empirical classes, as opposed to universal features shared across classes.

\section{The normalized Laplacian and its spectrum}
We model a network as  a  graph $\Gamma$ with $N$
vertices or nodes.  Two vertices $i,j
\in\Gamma$ are called neighbors, $i\sim j$, when they are connected by
an edge of $\Gamma$. For a vertex $i \in
\Gamma$, let $n_i$ be its degree, that is, the number of its
neighbors. For functions $v$ from the vertices of $\Gamma$ to
$\mathbb{R}$, we define the (normalized) Laplacian as
\bel{3}
\Delta v(i):=  v(i) -\frac{1}{n_i} \sum_{j, j \sim i}v(j) .
\end{equation}
This is different from the algebraic graph Laplacian
  usually   studied in the graph theoretical literature, see e.g. \cite{Bol}, but
  equivalent to the Laplacian investigated in \cite{Chung}. This
  normalized Laplacian is, for example, the operator underlying random
  walks on graphs, and in contrast to the algebraic Laplacian, it
  naturally incorporates a conservation law. \\
The spectrum, that is, the collection of eigenvalues of $\Delta$,
yields 
important invariants of the underlying graph $\Gamma$ that
incorporate its qualitative properties, for example, how difficult it
is to decompose the graph, or how different it is from a bipartite
graph, that is, one with two types of vertices where connections are
only permitted between vertices of different type (see
\cite{Chung}). Also, the spectrum controls the behavior of dynamical
processes supported by the network (see \cite{JJ1,J1}). One can essentially recover the graph from its spectrum
(for
a heuristic algorithm, see \cite{IpMi}), up
to isospectral graphs. The latter are known to exist, but are
relatively rare and qualitatively quite similar in most respects. \\
The multiplicity $m_1$ of the eigenvalue 1 of $\Delta$ is particularly
significant. $m_1$ is the number of linearly independent solutions of 
$\Delta v(i)=v(i)$ for all $i$, that is, of
\bel{4}
\sum_{j, j \sim i}v(j)=0 \text{ for all }i.
\qe
(Equivalently, $m_1$ is the dimension of the kernel of the adjacency
matrix of $\Gamma$.) -- Such functions can be created by node duplication: Take any node $i_0
\in \Gamma$ and form a new graph $\Gamma_0$ by adding a new node $j_0$
to $\Gamma$ and connecting it to all neighbors of $i_0$. Thus, in
$\Gamma_0$, $i_0$ and $j_0$ have the same neighbors. A solution $v$ of
\rf{4} on $\Gamma_0$ then is obtained by putting $v(i_0)=1, v(j_0)=-1$
and $v(i)=0$ for all other nodes $i$. In other words, node duplication
increases $m_1$ by 1. For this reason, it constitutes an important
invariant for our investigation of protein-protein interaction
networks. -- In a similar vein, doubling an edge that connects vertices
$p_1,p_2$ produces the eigenvalues $\lambda=1\pm
\frac{1}{\sqrt{n_{p_1}n_{p_2}}}$ which are symmetric about 1, and close
to 1 when the degrees are sufficiently large. -- Also, if we duplicate a particular node $m$ times, then the number of specific motifs containing that node will grow like $m \choose 2$; again that then is something that can easily be detected in given network data. \\

\begin{figure}[h]
\begin{minipage}{\textwidth} 
\begin{center}
\includegraphics[scale=0.5]{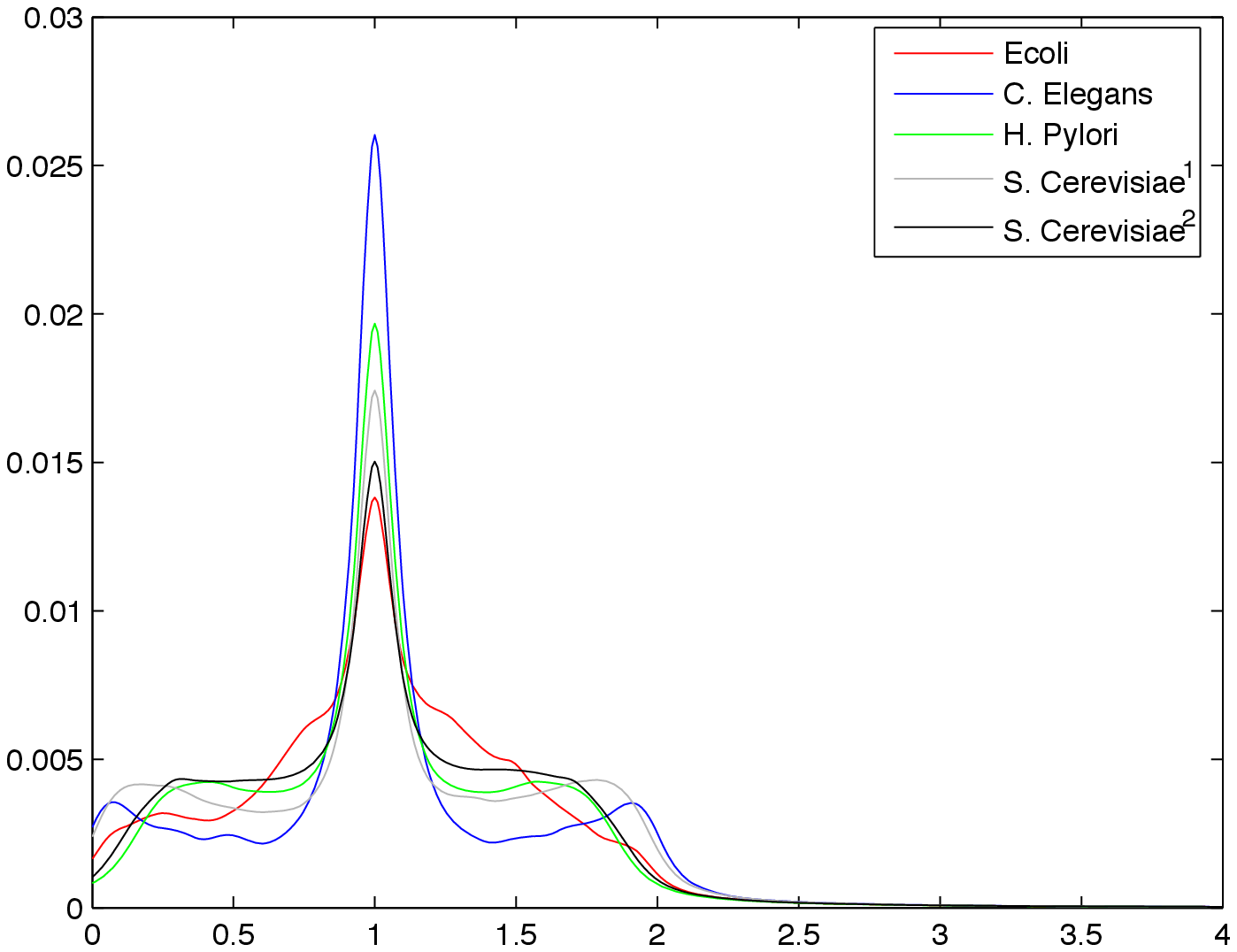}
\end{center}
\caption{}
\label{Laplacian_spectral_plots}
\end{minipage}
\end{figure}

\section{Spectral plot and structural analysis of protein-protein
  interaction networks}
In spite of their rather wide range of  sizes and in spite of possible
statistical fluctuations affecting the acccuracy of the underlying data, the spectral plots of
the different PPINs\footnote{See data source for details.}
 share a particular pattern
(Fig.\ref{Laplacian_spectral_plots}; the spectral density is given as a
sum of Lorentz distributions, $\rho(\lambda)=\sum_{k=1}^{N-1}
\frac{\gamma}{(\lambda_k -\lambda )^2+\gamma^2}$ with width
$\gamma=.08$ where $\lambda_1, \dots , \lambda_{N-1}$ are the nonzero eigenvalues). The most prominent feature is
the sharp peak around the eigenvalue 1.\footnote{A high multiplicity
  of eigenvalue 1 has also been observed in other networks, like the
  Internet \cite{VukadinovicEtAl2002}.} Also, the large degree of
symmetry around 1 is noteworthy. -- As a control, the various
important structural parameters also have typical ranges; examples
are, $N$ being the size of the network: 
Maximum degree $< \frac{N}{10}$, 
$1.56N <$ Number of edges $< 1.97N$, 
$0.307N < m_1 < 0.445N$, 
$0.015<$ Transitivity (relative frequency of vertex triangles) $<
0.028$.\\
In particular, the multiplicity $m_1$ of the eigenvalue $1$ and
the transitivity are much larger than in random graphs
of Erd\"os-R\'enyi type with a similar number of vertices and
edges. Similar observations hold for small motifs, that is, subgraphs
of a particular type, like cyclic chains of 4 vertices or structures
where 3 vertices do not have direct connections, but are connected
each to a central 4th vertex (data not shown).
\medskip

\section{Model and network reconstruction}
On the basis of the spectral analysis, a constructive model for the
evolution of a PPIN network can be proposed. The criterion is that the
model reproduce all the essential spectral features of the data
class. Our constructive model for PPINs is inspired by general evolutionary
considerations. The basic  evolutionary processes  for growth and
evolution of PPINs are duplication of  protein (nodes) and mutation of  connections (edges).\\
Instead of cross links between the old protein and its duplicated copy
-- which would produce too small values for the transitivity --,    a   low probability preference for 2nd order neighbors as
recipients of new connections is assumed. New connections
with other proteins then occur with a different probability. Since in link dynamics, attachment occurs
preferentially towards partners of high connectivity
\cite{BergEtAl2004},    some preferential
attachment to proteins with higher connections is included. In contrast, deletion
is random with a uniform probability. 

Since genome evolution analysis
\cite{Wagner2001,HuynenBork1998} on one hand supports the idea that
the divergence of duplicated genes takes place shortly after the
duplication, but on the other hand only indirect evidence is available
for rapid functional divergence after gene duplication
\cite{Wagner2001}, we have considered two different mutation processes:

\begin{enumerate}
\item A random deletion process that is independent of the duplication
  process  occurs uniformly with probability  $\delta$, and two different kind of addition processes with preference towards a partner with high degree.

\begin{enumerate}
\item Connection with protein $i$ at distance 2 with probability $\frac{d_i}{\sum_i d_i}\alpha_1$ , where $d_i$ is the degree of protein $i$ and $\alpha_1$ is a parameter. 

\item Connection with another protein $i$ (that could even be in
  another component) with probability $\frac{d_i}{\sum_i d_i}\alpha_2$, with a parameter $\alpha_2$.
\end{enumerate}

\item A   deletion with probability $\delta^\prime$  that
  occurs for $\frac{1}{3}$ of the duplications  and shortly after
  such a duplication. This process operates by elimination of one of
  the two interactions in each redundant interaction pair of two
  duplicate proteins with equal probability. For simplicity, there is no
  addition for this mutation process. 
\end{enumerate}

To make the duplication process independent of the first mutation
process and to make the duplication rate lower than the mutation rate,
duplication  occurs with probability $P_{\hbox{\textit{dup}}}$ and
with a preference that is the inverse of the square-root of the degree of the protein.

A component of the network can grow by duplication of proteins within
that  component or attachment of other components or isolated proteins. 

Here, we have neglected isolated proteins, but the model can be
readily extended by attachment of isolated proteins  with some
probability $P_{\hbox{\textit{add}}}$. One might also include a
mechanism for cross
link connections between duplicate protein pairs with some probability
$P_{\hbox{\textit{CLink}}}$, but the same effect can be achieved by tuning the other parameters.

The algorithm starts with a small seed network of two linked
proteins. 
The growth procedure is run until  the giant component reaches our
desired network size. 100 repetitions are performed with parameter
values 
$P_{\hbox{\textit{dup}}} = 0.15$, 
$\delta^\prime = 0.7$, 
$\delta = 0.00025$, 
$\alpha_1 = 0.00008$, 
$\alpha_2 = 0.0002$, 
$P_{\hbox{\textit{add}}} = 0.025$, 
$P_{\hbox{\textit{CLink}}} = 0.008$. 

The structural properties of the resulting giant component (size
$\thickapprox$ 500) are: 
Maximum degree  $\thickapprox 43.69$, 
Number of edges  $\thickapprox 712.97$, 
$m_1 \thickapprox 161.07$, 
Transitivity  $\thickapprox 0.02793$.\\

\begin{figure}[h]
\begin{minipage}{\textwidth} 
\begin{center}
\includegraphics[scale=.5]{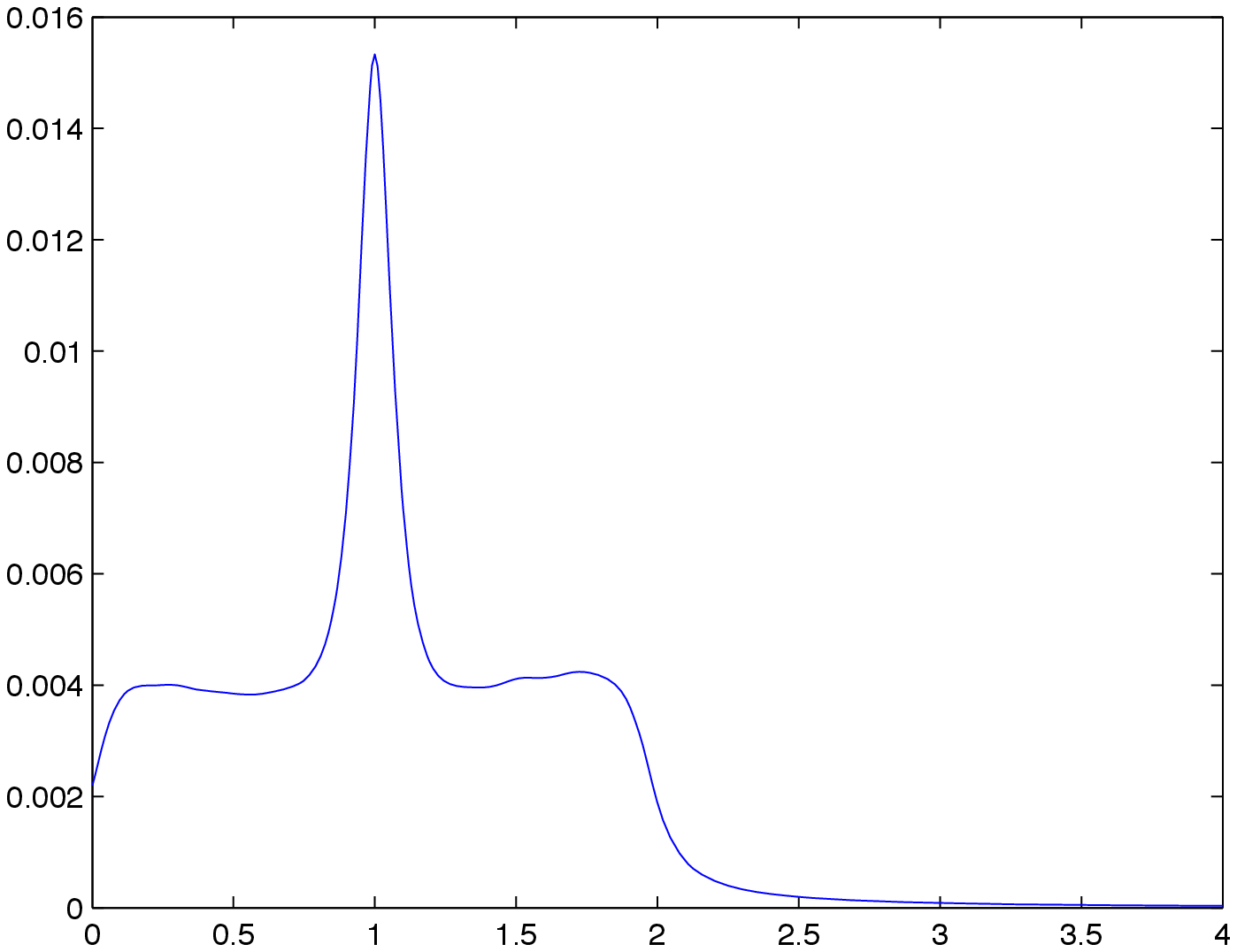}
\end{center}
\caption{} 
\label{prot_MixedAlgo2}
\end{minipage}
\end{figure}

Thus, the spectral plot (Fig.\ref{prot_MixedAlgo2}) and the structural
properties of the giant component of the simulated network match the
real PPIN data closely.\footnote{The spectrum of the Laplacian is
  always confined between 0 and 2. This is not quite exhibited by our
  spectral plots, due to the positive width of the kernel employed in
  our visualization.}\\\\

\begin{figure}[h]
\begin{center}

\includegraphics[width=.33\textwidth]{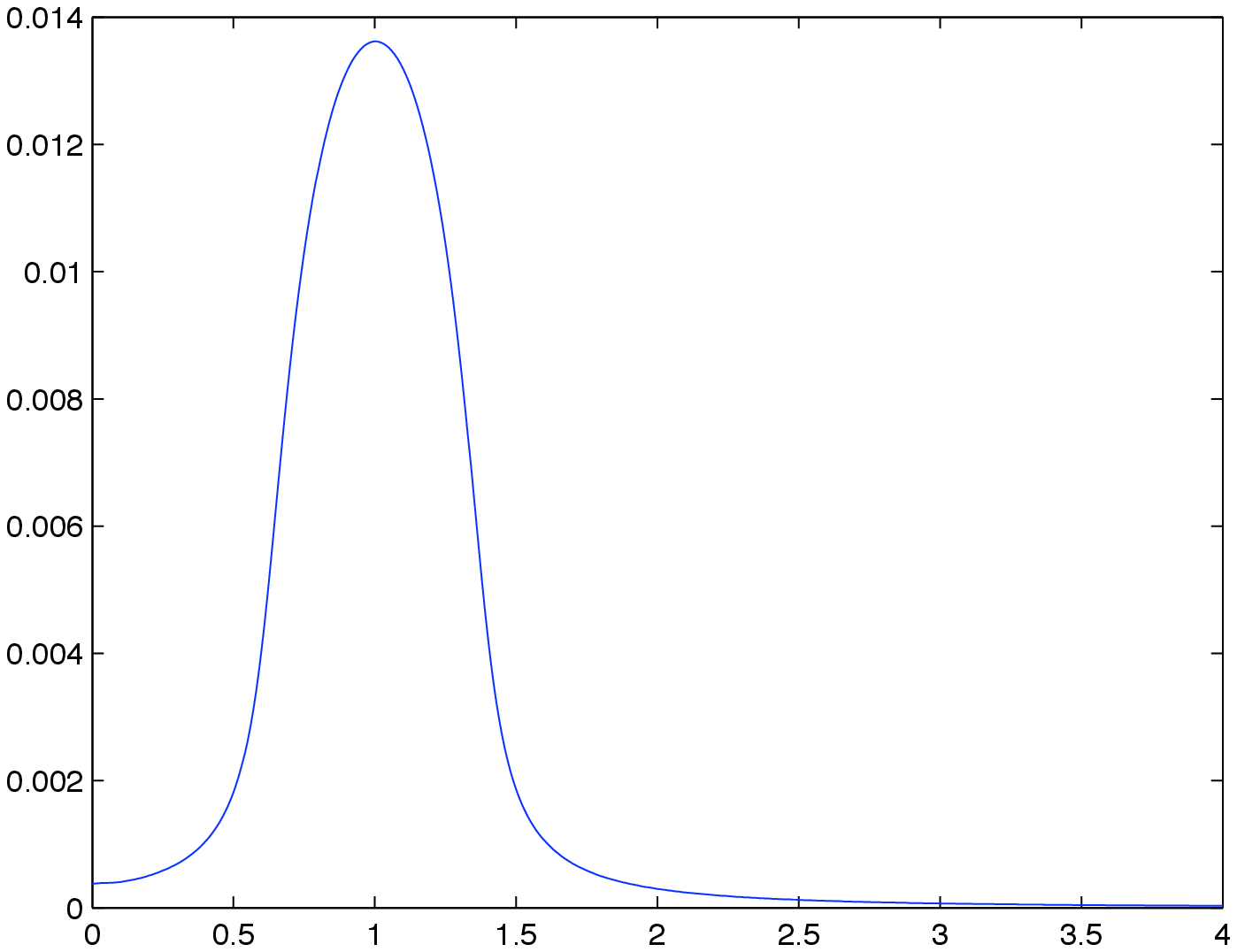}\includegraphics[width=.33\textwidth]{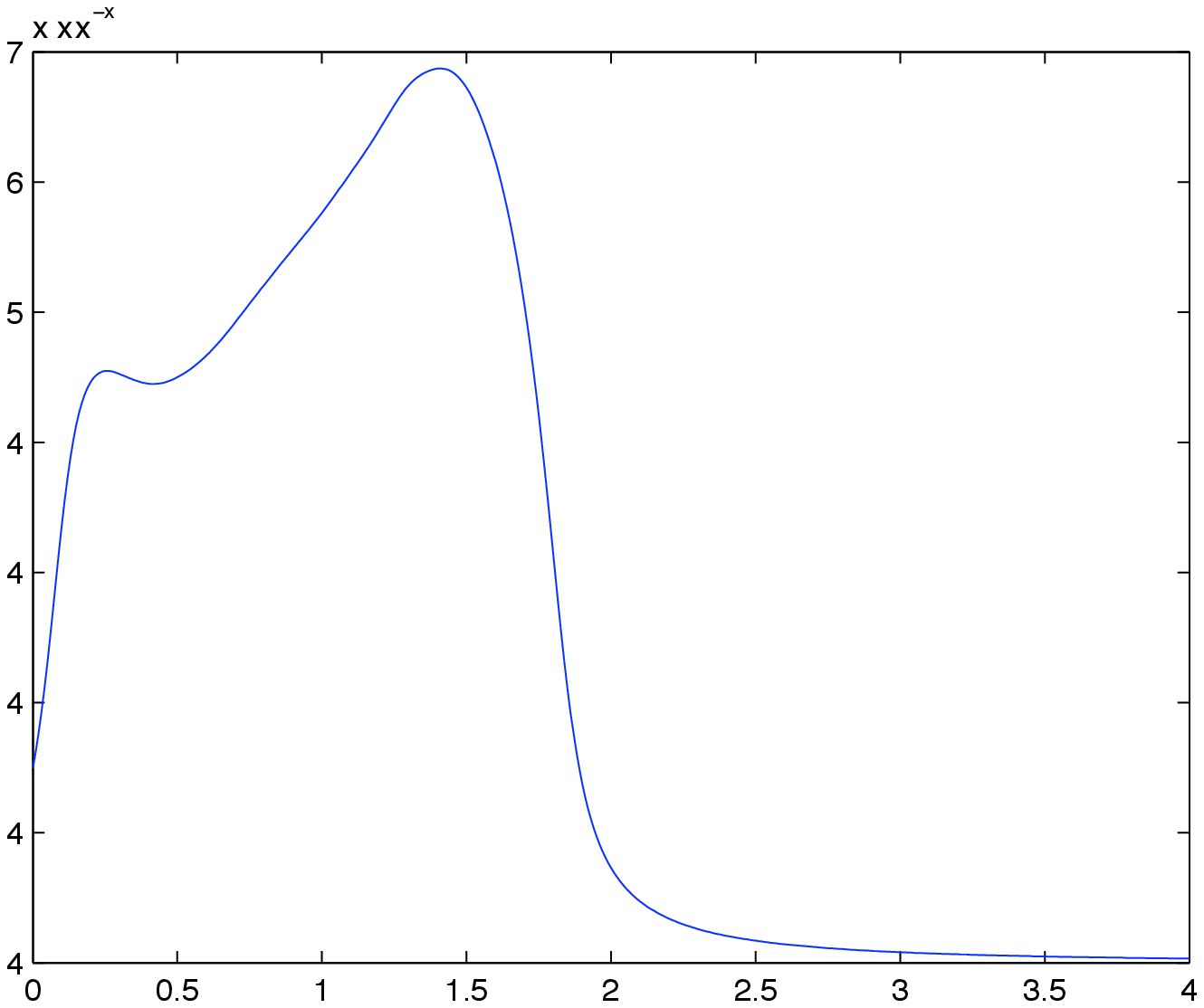}\includegraphics[width=.33\textwidth]{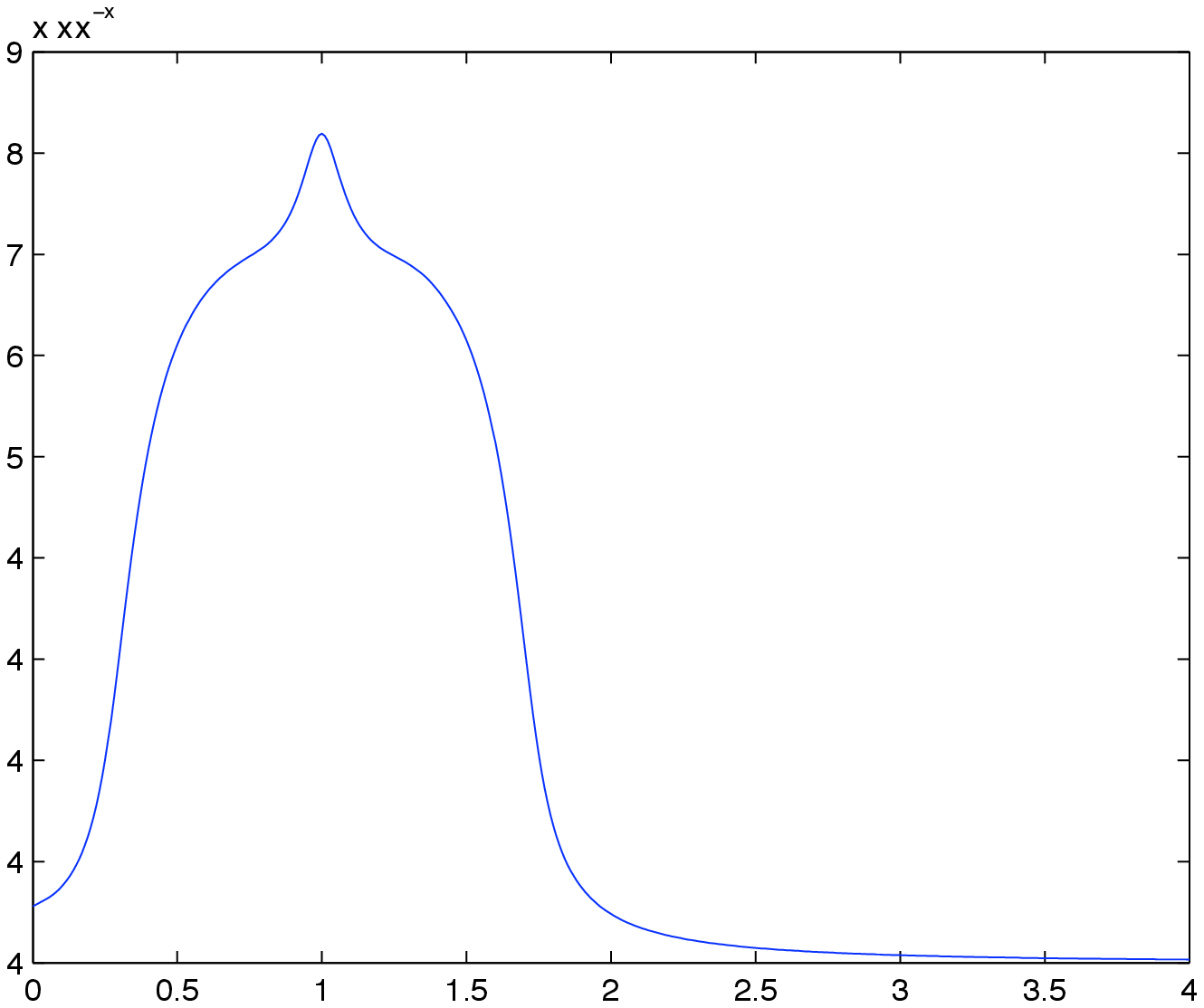}\\

(a)\hspace{.3\textwidth}(b)\hspace{.3\textwidth}(c)

\end{center}
\caption{Spectral plots of (a) a  random network by the Erd\"{o}s-- R\'{e}nyi model \cite{ER} with $p = 0.05$, (a) a small-world network by the Watts--Strogatz model \cite{WS} (rewiring a regular ring lattice of average degree 4 with rewiring probability $0.3$), (c) a scale-free network by the Barab\'{a}si--Albert model \cite{BA} ($m_0 = 5 \text{ and } m = 3 $). {\it Size of all networks is $500$. All figures are plotted with $100$ realizations.}}
\label{general_models}
\end{figure}

A comparison with generic network construction algorithms shows that
they necessarily important structural properties that are
characteristic for PPIN networks and distinguish them from networks
from other biological or nonbiological realms. Prominent examples of
such generic schemes are a regular network, the random
network of Erd\"os-R\'enyi\cite{ER}, the scalefree network construction by
preferential attachment of Barab\'asi-Albert\cite{BA}, and the small-world
network by random rewiring of a regular network of
Watts-Strogatz\cite{WS}. Spectral plots of such networks, with the
corresponding parameters adjusted to match the ones found for PPIN
networks and constructed by the same scheme as in our algorithm, are obviously qualitatively different from the ones for the
real data and our reconstructed network (see Fig.\ref{general_models}). This indicates that our
spectral analysis uncovers features that are specific for PPIN
networks. \\

Other previous reconstruction schemes (\cite{IspolatovEtAl2005,Pastor-SatorrasEtAl2003,KimEtAl2002}) typically focus on certain
individual parameters in distinction to our emphasis on the entire
spectrum. Consequently, the spectral plots are also different (details
not shown). The model of
 \cite{VazquezEtAl2003} includes a  parameter $p$ that incorporates the probability
 of cross interactions between the old protein and its duplicated
 copy, for example resulting from self-interactions of the old one.
 A realistic value of  $p$ can then  be determined from the data in
 \cite{Wagner2001, Wagner2003} and is smaller than $0.018$. That upper
 bound is the value employed in \cite{VazquezEtAl2003}, but this
 scheme, for example, leads to too small a value for the transitivity of the giant
 cluster. Therefore, in our model we  assumed that, with some low
 probability, there is a preference for a protein to make new
 connection with its 2nd neighbors.

\section*{Data Sources}
The protein protein interaction data sets for  \textit{Saccharomyces
  cerevisiae}$^1$ (yeast) are from http://www.nd.edu/$\sim$networks/, used in
\cite{JeongEtAl2001} [download date: 17th September, 2004]. The ones for
\textit{Escherichia coli} as used in \cite{ButlandEtAl2005},
\textit{Caenorhabditis elegans, Helicobacter pylori} and, as a check,  a
second data set for \textit{Saccharomyces cerevisiae}$^2$ are taken from
http://www.cosin.org/ [download date: 25th September, 2005]. Note that
these two data sets on the same cell are quite different. This
indicates the robustness of our method in view of possibly significant
statistical fluctuations of the data employed. -- Our
analysis has been always performed on the giant components of these
networks so as to work with connected graphs, and
we have neglected the many small
components and isolated proteins.

\bibliographystyle{plain}

\end{document}